\documentclass[final,5p,times,twocolumn]{elsarticle}

\usepackage{amssymb}
\usepackage{lipsum}
\usepackage[version=3]{mhchem} 
\usepackage{multirow}
\usepackage{subcaption}
\usepackage{multirow}
\usepackage{hyperref}
\usepackage{color}
\usepackage[export]{adjustbox}
\usepackage{dblfloatfix}

\graphicspath{ {./images/} }

\begin{document}
	
	\begin{frontmatter}
		
		

		\title{HF Etching and Silanization: Evidence for the Role of Surface Hydroxyl Groups in Silicon Nitride Resonator Loss}

		\author[first]{Ariane Giesriegl}
		\author[first]{Nicola Cavalleri}
		\author[first]{Robert West}
		\author[second]{Antonius Armanious}
		\author[third]{Markus Sauer}
		\author[fourth]{Thomas Schachinger}
		\author[first]{Silvan Schmid}
		\affiliation[first]{organization={Institute of Sensor and Actuator Systems, TU Wien},
			addressline={Gusshausstrasse 27-29}, 
			postcode={1040}, 
			state={Vienna},
			country={Austria}}
		\affiliation[second]{organization={Laboratory for Solid-State Physics, ETH Zürich},
			addressline={Otto-Stern-Weg 1}, 
			postcode={8093}, 
			state={Zürich},
			country={Switzerland}}
		\affiliation[third]{organization={Analytical Instrument Center, TU Wien},
			addressline={Getreidemarkt 9}, 
			postcode={1060}, 
			state={Vienna},
			country={Austria}}
		\affiliation[fourth]{organization={Service Unit of University Service Centre for Transmission Electron Microscopy, TU Wien},
			addressline={Wiedner Hauptstraße 8}, 
			postcode={1040}, 
			state={Vienna},
			country={Austria}}
		
		\begin{abstract}
			Silicon nitride (\ce{SiN_x}) nanomechanical resonators are central to sensing, quantum technologies, and fundamental physics experiments due to their exceptional mechanical quality factors (Q). However, as resonator thickness approaches the nano-scale, surface-related dissipation limits performance. Here, we investigate the role of surface chemistry in low-stress Si-rich SiNx membranes through a combination of hydrofluoric acid (HF) etching and trimethylchlorosilane (TMCS) silanization, correlated with surface characterization and mechanical measurements. Preliminary analysis by TEM-EELS, XPS, RBS/ERDA, and XRR reveals a native oxide surface layer (1–2\,nm). Surface modification by HF and TMCS was subsequently evaluated using XPS, photothermal FTIR, contact-angle measurements, and intrinsic quality factor ($Q_{int}$) characterization. While HF etching effectively removes the native oxide and TMCS introduces hydrophobic \ce{Si-(CH3)3} termination, neither oxide thickness nor surface energy correlates with mechanical dissipation. TMCS treatments produce the largest enhancements, increasing $Q_{int}$ by up to 50\%, whereas HF etching alone yields lower gains of 20–25\%. These findings suggest surface hydroxyl groups as a key contributor to energy loss in \ce{SiN_x} resonators and demonstrate that chemical functionalization can substantially suppress surface dissipation.
		\end{abstract}
		
		
		
		
	\end{frontmatter}
	
	
	
	
	\section{Introduction}
	\label{introduction}
	
	\begin{table*}[b!]
		\begin{center}
			\caption{Characterization results of the plain samples without treatment. XPS signals have been taken at an electron take-off angle of 45°. Measurement details can be found in the methods section.}
			\label{tab:prelim}
			\scalebox{1}{
				\begin{tabular}{lccc} 
					\hline
					&	Silicon		&		100\,MPa	&	200\,MPa		 \\
					\hline
					Intrinsic Q-factor $Q_{int}$			&				&	3379 $\pm$ 478	& 3060 $\pm$ 346	 \\
					\ce{SiN_x} characteristics				&				&					&					\\
					\hspace{.5cm} N/Si ratio 				&				& 					&					 \\
					\hspace{1cm} XPS (bulk)					&				& 		1.01		&      1.09			 \\
					\hspace{1cm} ERDA    					&				& 		0.94		&  		1			 \\
					\hspace{1cm} RBS						&				& 	0.95 $\pm$ 0.05	&  	1.09 $\pm$ 0.05	 \\
					\hspace{.5cm}Thickness [nm]				&				& 					&					 \\
					\hspace{1cm} TEM						& 		-		&		45 $\pm$ 2	& 	   46 $\pm$ 2    \\
					\hspace{1cm} XRR						& 		-		& 		47.6   		& 		49.4		 \\
					\hspace{.5cm}Density [$g/cm^3$] (XRR)	& 				& 		2.9			&		3.0			 \\
					Surface characteristics					&				&					&					\\
					\hspace{.5cm}Contact Angle [°]			& 54 $\pm$ 2 	& 	35 $\pm$ 5 		& 	50 $\pm$ 4		 \\
					\hspace{.5cm}RMS roughness [pm]			& 			63 	& 		536 		& 			448		 \\
					\hspace{.5cm}O/Si ratio (XPS)		    & 		0.57	&   	0.68		&      0.70			 \\
					\hspace{.5cm}C/Si ratio (XPS)		    & 		0.22    &   	0.20		&      0.17			 \\
					\hspace{.5cm}Oxide thickness [nm] (XPS)	&1.78 $\pm$ 0.1	&  1.68 $\pm$ 0.03  &   1.74 $\pm$ 0.04	 \\
					\hline
				\end{tabular}
			}
		\end{center}
	\end{table*}
	Silicon nitride is a widely used material for nanomechanical resonators due to its favorable optical and mechanical properties, such as broadband transparency and high intrinsic tensile stress \cite{Kanellopulos2024,Sementilli2021}. High tensile stress is particularly advantageous as it generates a quasi-lossless potential that dilutes intrinsic losses in the resonators, resulting in Q-factor values of several million. \cite{schmid2011damping}
	High Q factors lead to higher force sensitivity and coherence, which are interesting for several research areas, such as quantum mechanic investigations at room temperature,\cite{Norte2016} magnetic resonance force microscopy with single nuclear spin sensitivity\cite{Poggio2010,Kosata2020,Haelg2021} and dark matter as well as gravitational wave detection.\cite{Page2021,Manley2021,Carney2021}
	
	Various strategies have been employed to further enhance the Q-factor, notably through optimized device geometries. For instance, the introduction of phononic crystal patterns can suppress clamping losses and minimize bending losses via so-called soft clamping \cite{tsaturyan2017ultracoherent,ghadimi2018elastic}. Improving Q further requires a better understanding of the remaining damping mechanisms and losses, particularly surface losses, since intrinsic dissipation in \ce{SiN_x} resonators has been shown to be surface-bound.\cite{Villanueva2014} In this work, the focus is set on investigating said surface losses.
	
	Thin resonators, in the range of tens of nanometers in thickness, are particularly useful to study surface losses, as surface effects dominate in structures with a high surface-to-volume ratio.\cite{Villanueva2014} Previous studies have highlighted the deteriorating effect of free surface charges and ions, defects states (two-level systems) and surface oxide layers on the Q factor in micro-mechanical resonators.\cite{Schmid2014,Ono2005,Heritier2021,Luhmann2017}
	
	Luhmann et al. demonstrated that increased oxide thickness, induced by oxygen plasma treatments, correlates with reduced tensile stress and lower intrinsic Q-factor \cite{Luhmann2017}. Conversely, Tao et al. showed that hydrofluoric acid (HF) etching enhances the Q factor in silicon resonators and that subsequent chemical protection helps stabilize this improvement. They suggest that dissipation arises from the amorphous structure of the oxide layer rather than its chemical composition \cite{Tao2015}. Henry et al. further reported that Q in silicon is maximized when resonators are terminated with a methyl monolayer directly grafted onto the surface, attributing this to a lower density of electronic defects and increased chemical stability \cite{Henry2011}. 
	
	Several sources of intrinsic dissipation — such as stress gradients between \ce{SiN_x} and native \ce{SiO_2}, defects within the amorphous interface layer (e.g., two-level systems), and surface adsorbates like adventitious carbon or water — can thus be mitigated by removing the native oxide and chemically tailoring the surface \cite{Luhmann2017,Tao2015,Heritier2021,Henry2011,Sementilli2021}.
	
	Surface silanization, i.e. the attachment of organosilyl groups onto a surface, has been studied on silicon or silica and several methods have been published in literature.\cite{Kinkel1984,Szkop2018,Maharanwar2017,Rezayi2015,Tao2015,Fadeev2000} Silanization needs surface hydroxyl groups to bind to the surface, which are naturally present on native oxide layers.\cite{Liu2016,Coffinier2022}  Raider et al. measured the growth of the native oxide layer on HF etched \ce{Si3N4}, and found that it already is $\approx 0.3\ nm$ on a freshly etched film, and that it grows in the course of 1 month to be $\approx 1\ nm$ thick, in terms of \ce{SiO_2} film thickness. They claim that the oxidation product is more likely a graded oxynitride layer, which is more oxygen-rich at the surface, a conclusion supported by other work. \cite{Raider1976,Wurzbach1983,Greil1991} 
	Lamagna et al. have determined similar amounts of surface hydroxyl and amine groups on pristine crystalline $\beta$-\ce{Si3N4}(0001) surfaces; around $3\text{-}4\ OH/nm^2$, which is slightly lower than the average -OH density on silica, which is $5\text{-}6\ OH/nm^2$.\cite{Lamagna2012} After the formation of a native oxide layer, the balance is shifted towards surface hydroxyl groups. \cite{Pezotti2024}
	However, surface hydroxyl groups may be scarce on freshly HF-etched \ce{SiN_x} surfaces. Some works reported the concentration of surface functional groups after HF etching of \ce{Si3N4}, which is mostly \ce{Si-F}, then \ce{Si-OH}, and some \ce{Si-NH2}.\cite{Brunet2017} They also claim that silanization leads to good film formation on silicon nitride surfaces, even after removal of the native oxide layer. \cite{Liu2016} 
	
	In this work, we employed a surface treatment strategy involving HF etching followed by wet-chemistry silanization with trimethylchlorosilane (TMCS), both under ambient and inert conditions. Additionally, we added a treatment with silanization alone, without HF etching, in both the liquid and gaseous phases. The aim of these treatments is partly the removal of the native oxide layer and the subsequent attempt to inhibit further oxidation of the surface by protecting it with silanes, which also modify the surface functional groups. Treatments were performed on low-stress silicon nitride films with varying silicon-to-nitrogen ratios. We observed a significant increase in Q for these samples, which is attributed less to the removal of the native oxide layer, but rather to the changes in the surface functional groups, which may change its interaction with adsorbates from the atmosphere.    
	
	To characterize the silicon nitride films, we performed compositional analyses, including Transmission Electron Microscopy with Electron Energy Loss Spectroscopy (TEM-EELS), X-Ray Photoelectron Spectroscopy (XPS), Atomic Force Microscopy (AFM), Ion Beam Analysis (including Rutherford Backscattering, RBS; and Elastic Recoil Detection Analysis, ERDA) and X-Ray Reflectivity (XRR). After surface treatments, the samples were further analyzed using Nanomechanical Photothermal Fourier-Transform Infrared Spectroscopy (NEMS-FTIR), Contact Angle (CA) analysis, XPS, and Q-factor measurements, which is the main indicator for reduced losses. The measurement of the Q-factor, more specifically the intrinsic Q-factor ($Q_{int}$), is independent of stress.
	
	\begin{figure*}[ht!]
		\centering
		\begin{subfigure}[b]{0.33\textwidth}
			\centering
			\includegraphics[width=\textwidth]{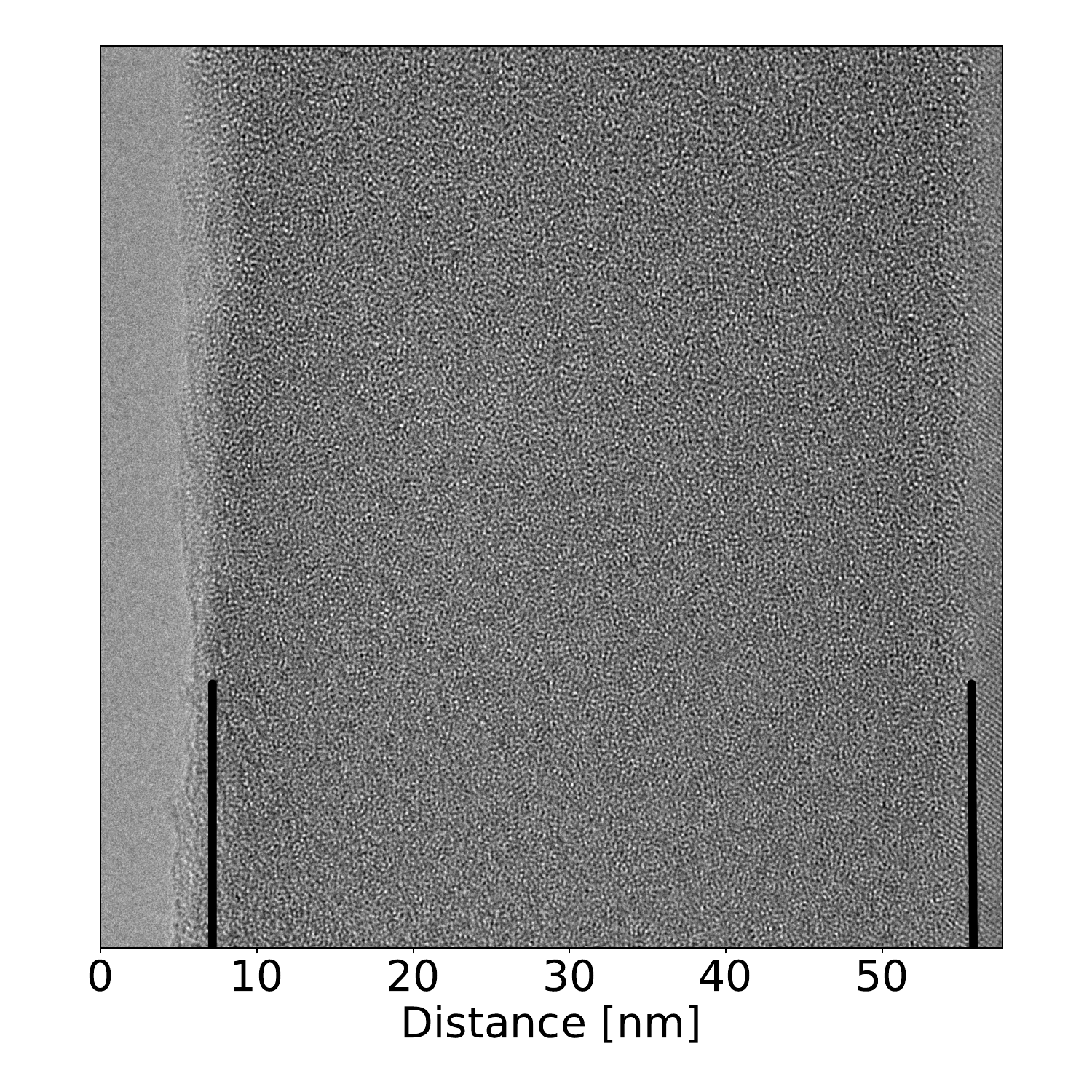}	
			\caption{}
			\label{fig:200MPaTEM}
		\end{subfigure}%
		\begin{subfigure}[b]{0.33\textwidth}
			\centering
			\includegraphics[width=\textwidth]{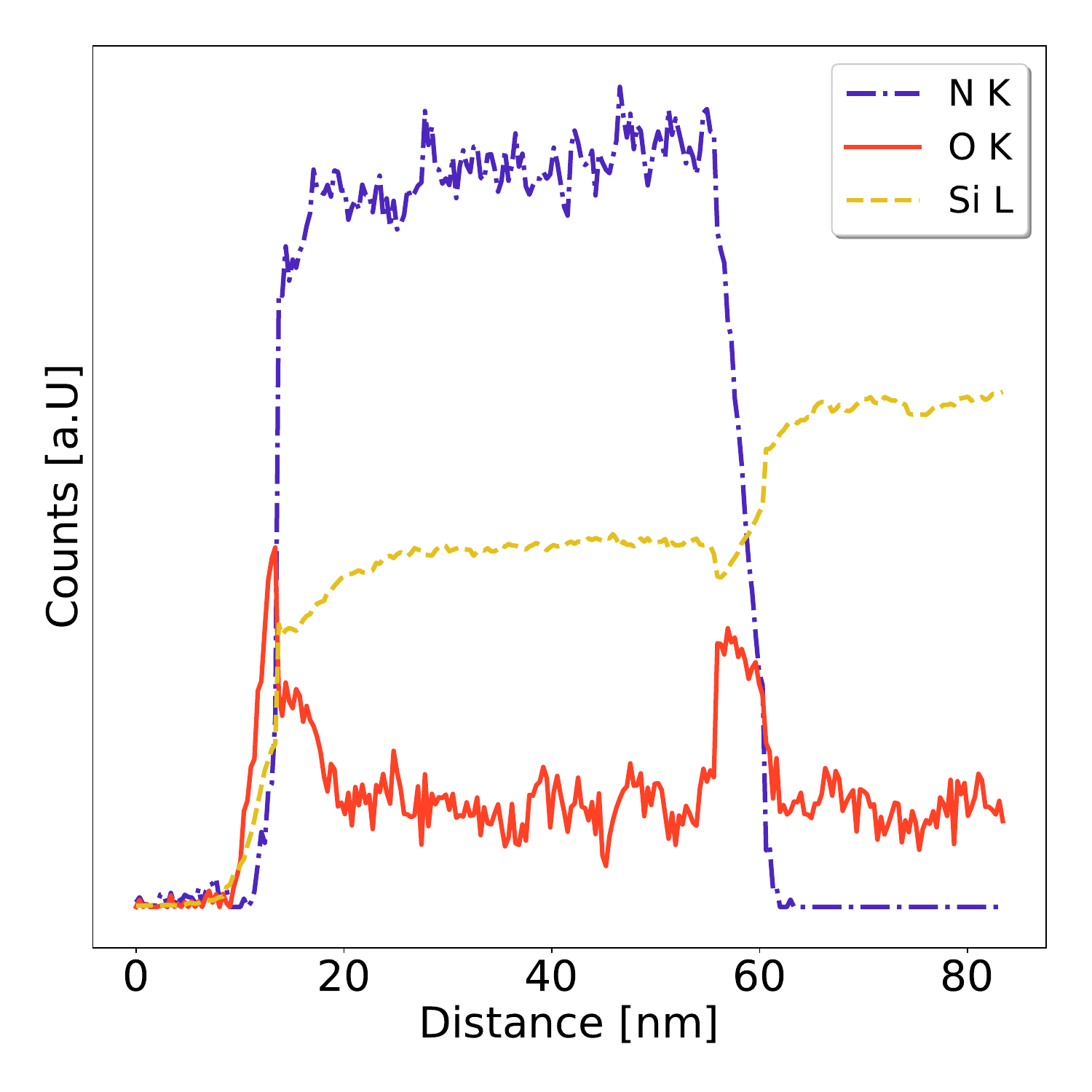}
			\caption{}
			\label{fig:200MPaEELS}
		\end{subfigure}%
		\begin{subfigure}[b]{0.33\textwidth}
			\centering
			\includegraphics[width=\textwidth]{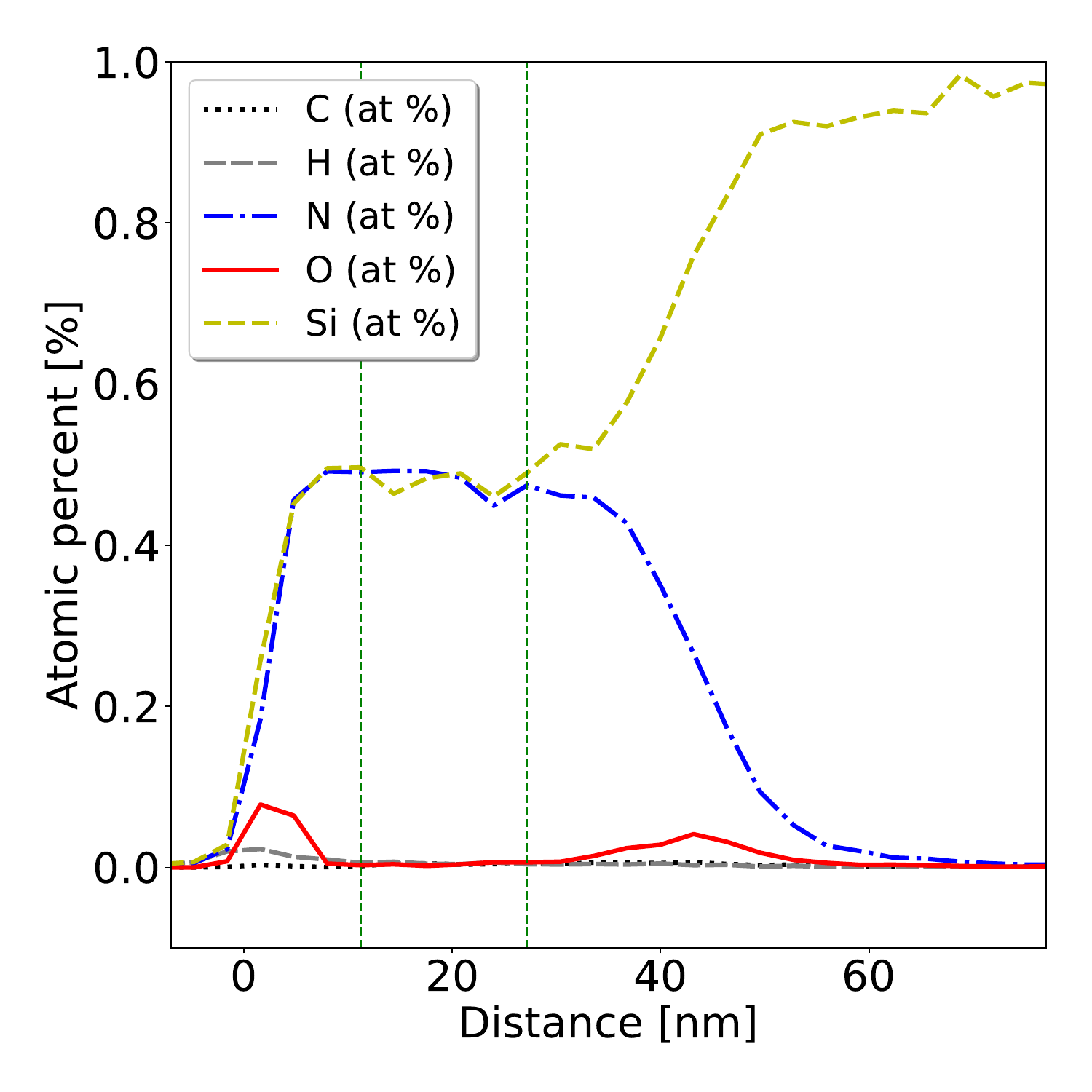}
			\caption{}
			\label{fig:200MPaERDA}
		\end{subfigure}%
		\caption{Three depth profiling methods for the 200\,MPa sample. a) Cross-sectional TEM image showing the different layers; b) EELS line scan analysis across the multilayers; c) ERDA depth profile; vertical lines indicate the range in which the elemental ratios were evaluated.}
		\label{fig:crosssection}
	\end{figure*}
	
	\section{Results and discussion}
	\subsection{Preliminary characterization}
	We used two different Si-rich silicon nitride samples, with their tensile stress set mainly by their nitrogen-to-silicon ratio (N/Si).\cite{Kanellopulos2024} The nominal stress of the samples are 100\,MPa and 200\,MPa, with which these samples will be referred to in this work. A summary of the preliminary characterizations is show in \autoref{tab:prelim}:
	
	The main figure of merit in this work is the intrinsic Q-factor ($Q_{int}$), which is derived from the measured Q according to \autoref{Eq:Qint} and eliminates the effect of stress-induced dissipation dilution, thereby isolating the intrinsic material losses for further comparison.\cite{Schmid2023} As can be seen, the $Q_{int}$ is around 3000 for both samples, though a bit higher in the 100\,MPa sample. This could be because we use the same Young's modulus $E$ value for both samples. 
	
	The N/Si ratio has been determined via three methods: XPS, RBS, and ERDA. The latter two are powerful ion beam analysis techniques used to characterize the composition and depth profiles of materials. It combines backscattering and recoil detection, enabling precise quantification of both heavy and light elements, including hydrogen, in thin films and multilayered structures. 
	
	XPS, RBS, and ERDA results show that the 100\,MPa sample has a lower N/Si ratio than the 200\,MPa sample, consistent with tensile stress being correlated with the N/Si ratio.\cite{Kanellopulos2024,TempleBoyer1998}
	
	The thickness of the \ce{SiN_x} layer has been determined via TEM images, in which more than 10 positions were measured to obtain a good average, and XRR, which gives the thickness as the result of a fit. The results show that the measured thickness is slightly below the nominal thickness of 50\,nm.
	
	The density has been determined via XRR, and show that the 100\,MPa sample has a slightly lower density than the 200\,MPa samples, 2.9 and 3.0 $g/cm^3$, respectively.
	
	The surface characteristics focus on surface sensitive measurements, to analyze the native oxide layer. Both \ce{SiN_x} samples are compared to a plain silicon (100) wafer.
	
	The lowest Contact Angle (CA) was observed on the 100\,MPa sample, indicating that this surface has either a higher surface free energy and/or lower roughness than the other surfaces.\cite{Han2019} The root mean squared (RMS) roughness values from AFM measurements show that the roughness in the 100\,MPa sample is slightly higher than in the 200\,MPa sample, which lead to the conclusion that the lower CA is not due to the roughness but rather due to a higher surface free energy.
	
	The high surface free energy could be due to more hydroxyl or more amine groups on the surface. The other samples show a higher CA - silicon and 200\,MPa have a CA around 54° - which could mean that they have similar surface functional groups, i.e., more apolar surface groups than the 100\,MPa sample. Other factors that influence the CA include adsorbed water and atmospheric hydrocarbons, which are minimized by cleaning the samples with ethanol before measurement.\cite{Bryk2020} Additionally, several XPS measurements have been done on the samples, and they show that the adventitious carbon signal in relation to the silicon signal (C/Si ratio) on the 100\,MPa sample is higher than or similar to the 200\,MPa sample (\autoref{tab:prelim}).
	
	The O/Si ratios are determined via XPS at an angle of 45°. The values for 100\,MPa and 200\,MPa are similar, indicating that there is little variation in the oxygen content for both \ce{SiN_x} samples. The value is lower for plain silicon, which could be due to the fact, that on silicon there is \ce{SiO_2}, while on \ce{SiN_x} there is a \ce{SiO_xN_y} - both these layers have different passivation limitations.\cite{Bohling2015} 
	
	With the Si/O ratio and angle-resolved XPS measurements, we can determine the oxide thickness in XPS, in terms of \ce{SiO_2} film thickness (see \autoref{Eq:scipy}). This has been determined to be $\approx 1.7\ nm$ for the \ce{SiN_x} samples and $\approx 1.3\ nm$ for silicon.
	
	To understand surface-related losses, it is important to do a comprehensive characterization of the surface. Cross sections of the samples were prepared for TEM measurements. The image shown in \autoref{fig:200MPaTEM} reveals distinct layers: A heterogeneous oxide layer (left), amorphous \ce{SiN_x} (center), and crystalline silicon (right), with interfaces marked by black lines. The \ce{SiN_x} thickness was determined to be 46$\pm$1\,nm. EELS line-scan analysis (\autoref{fig:200MPaEELS}) across the multilayer structure reveals a thin ($\approx$1\,nm) \ce{SiO_x} overlayer, followed by a \ce{SiO_xN_y} interlayer and a bulk \ce{SiN_x} layer. The elevated oxygen signal in the \ce{SiN_x} and Si bulk is attributed to sample preparation: oxidation of the exposed surface occurred during the several-week interval between preparation and measurement. An additional oxide layer is evident at the \ce{SiN_x}/Si interface, coming from the native oxide of the silicon wafer substrate. The total \ce{SiN_x} layer thickness, including the oxynitride interlayer, was measured at $\approx$46\,nm, consistent with TEM analysis.
	
	ERDA results of the 200\,MPa sample (\autoref{fig:200MPaERDA}) show minor H and O signals at the surface, attributed to surface hydroxyl groups, adsorbed moisture, and native oxide. Bulk compositions consist primarily of Si and N, with trace concentrations of C, O, and Cl. Within the $\approx$11-27\,nm depth range (see \autoref{fig:200MPaERDA}), H/N ratios of 0.011 and 0.013 were measured for the 200\,MPa and 100\,MPa samples, respectively (see Supporting Information for additional data). A secondary rise in O signal occurs at the \ce{SiN_x}/Si interface, consistent with EELS observations.
	
	Comparable measurements for the 100\,MPa sample are shown in the Supporting Information.
	
	\begin{table*}[hb]
		\centering
		\caption{Oxide thickness determined from XPS measurements, in terms of \ce{SiO_2} thickness with a $\rho=1.5\ g/cm^3$.}
		\label{tab:XPS}
		\begin{tabular}{lcccc} 
			\hline
			&	untreated		&		HF			&	HF+TMCS(l) air	&	HF+TMCS(l) air	\\
			Sample								&					&	same day		&		1 month		&	6 months		\\
			\hline
			Silicon 							&					&					&					&					\\
			\hspace{.5cm}O/Si					&		0.66		&					&		0.43		&					\\
			\hspace{.5cm}F/Si					&		0.00		&					&		0.00		&					\\
			\hspace{.5cm}$d_{\ce{SiO_2}}$ [nm]	&	1.78$\pm$0.1	&					&	1.16$\pm$0.03	&					\\
			100\,MPa							&					&					&					&					\\
			\hspace{.5cm}O/Si					&		0.67		&		0.26		&		0.40		&		0.55		\\
			\hspace{.5cm}F/Si					&		0.00		&		0.04		&		0.03		&		0.03		\\
			\hspace{.5cm}$d_{\ce{SiO_2}}$ [nm]	&	1.68$\pm$0.03	&	0.7$\pm$0.02	&	1.03$\pm$0.02	&	1.4$\pm$0.04	\\
			200\,MPa							&					&					&					&					\\
			\hspace{.5cm}O/Si					&		0.72		&		0.25		&		0.42		&		0.54		\\
			\hspace{.5cm}F/Si					&		0.00		&		0.03		&		0.05		&		0.01		\\
			\hspace{.5cm}$d_{\ce{SiO_2}}$ [nm]	&	1.74$\pm$0.04	&	0.66$\pm$0.01	&	1.05$\pm$0.03	&	1.46$\pm$0.02	\\
			\hline
		\end{tabular}
	\end{table*}
	
	\subsection{Surface treatments}
	
	The samples were subjected to different treatments to study the losses on the surface by either removing the native oxide layer, modifying the surface functional groups or both. More specifically, the surface treatments included hydrofluoric acid (HF) etching, silanization with TMCS in the gas phase (TMCS(g)) or liquid phase (TMCS(l)), and combined HF + TMCS treatments in liquid performed in either ambient (HF+TMCS(l) air) or \ce{N2} (HF+TMCS(l) \ce{N2}) atmosphere.\\
	
	\begin{figure}[!t]
		\includegraphics[width=\columnwidth]{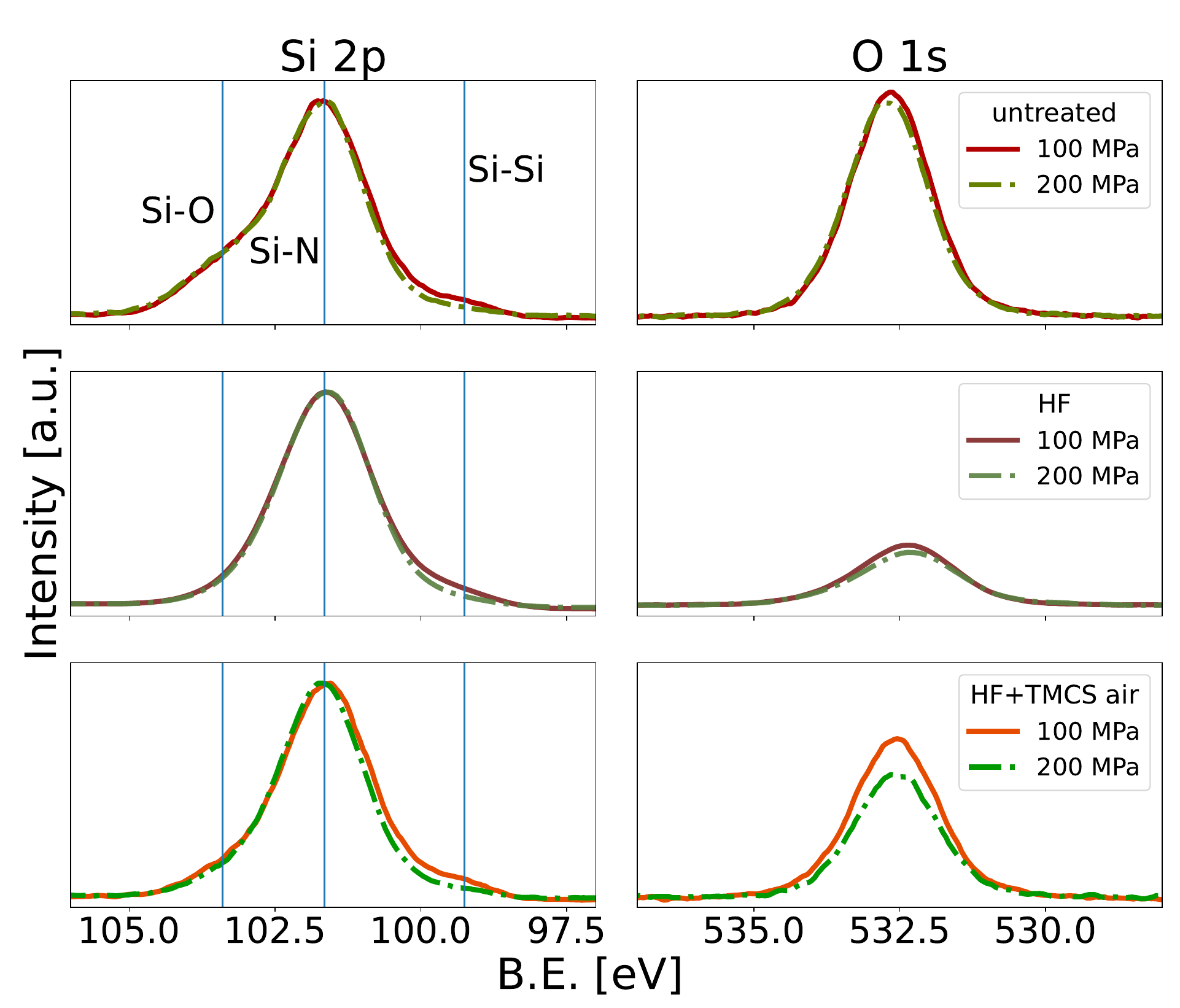}
		\caption{Left: Normalized Si\,2p signal measured at 45° with the Si-O, Si-N and Si-Si binding energies indicated as lines. Right: O\,1s signal normalized to the Si\,2p reference; y-axis limits are doubled to accommodate the larger signal intensity. The upper plots show the untreated 100 and 200 MPa samples, the middle row the HF only samples, while the lowest row the HF+TMCS air samples measured after 6 months.}
		\label{fig:XPS}
	\end{figure}
	
	\autoref{fig:XPS} shows XPS analysis at 45° for three treatment types: untreated (top), HF-etched same-day (middle), and HF+TMCS(l) air-exposed six months post-treatment (bottom). Both 100 and 200\,MPa samples have similar Si\,2p profiles (left), dominated by Si-N at 101.6 eV, some Si-O at higher B.E. (at 103.2 eV) and minor contributions from Si-Si at 99.2\,eV. The 100\,MPa sample shows a slightly higher Si-Si peak, indicating higher silicon content at lower stress. The O\,1s signal (right) shows comparable oxygen levels at both samples, primarily from the native oxynitride layer, with small contributions from surface hydroxyl groups and oxygen-containing airborne adsorbates.
	
	After HF etching, the Si-O peak is reduced for both samples, in accordance with the O\,1s signal being significantly lower, while there is no obvious change in the Si-N and Si-Si contributions to the peak. Some oxygen is still detected in the O\,1s spectrum, which either be due to quick oxidation of the surface, adsorbed moisture or some subsurface oxygen in the top of the \ce{SiN_x} layer.\cite{Liu2016}
	
	When the sample was treated with HF+TMCS(l) in air and measured six months later, an increase in the Si-O and O\,1s peaks was observed. This indicates that silanization does not inhibit surface oxidation on low-stress \ce{SiN_x}.
	
	The slow increase of the O/Si ratio over time can also be seen in \autoref{tab:XPS}. It is reduced to $\approx 35\%$ of the original value by HF etching, and then slowly increases for the HF+TMCS samples to $\approx 60\%$ and $\approx 80\%$ after one and six months, respectively. As a comparison, on pure silicon, the O/Si ratio is reduced to $\approx 65\%$ after one month.
	
	The table also shows that a small amount of fluoride remained on the surface after the HF etching. The amount was similar in both samples.
	
	\begin{figure}[!t]
		\includegraphics[width=\columnwidth]{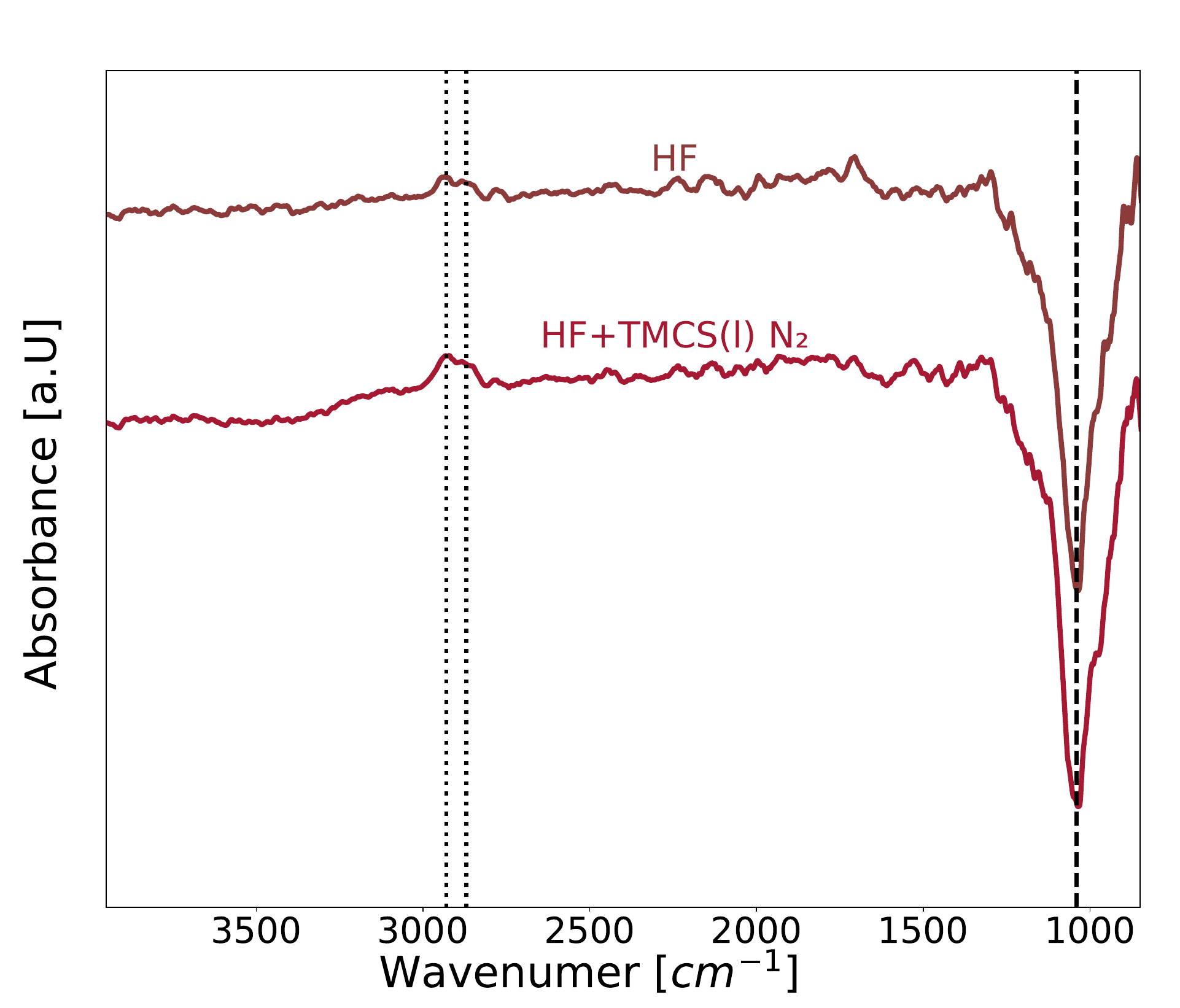}
		\caption{FTIR difference spectra (in reference to the untreated sample) of 100\,MPa samples: HF and HF+TMCS \ce{N2}. Dotted lines indicate the C-H vibrations, the dashed line indicates the negative Si-O-Si peak.}
		\label{fig:FTIR}
	\end{figure}
	
	The difference FTIR spectra from 800 to 3500 $cm^{-1}$ with the untreated sample spectrum subtracted from the treated sample spectrum are shown in \autoref{fig:FTIR}. The HF-treated sample without silanization clearly shows a negative peak at 1040 $cm^{-1}$ (dashed line), indicating the removal of the native silicon oxynitride surface layer.\cite{Giesriegl2019} The HF etched and silanized with TMCS in $N_2$ atmosphere sample shows a negative peak in similar size. Both spectra also show small peaks around 2800-2900 $cm^{-1}$ (dotted line), which come from the aliphatic C-H vibrations. They either stem from the silanization monolayer or from adventitious carbon.\cite{Lisovskyy2023,Laades2012,Rebib2008} The FTIR spectra were taken 1 week after the treatments, confirming that the native oxynitride layer on \ce{SiN_x} grows slowly.\cite{Raider1976} Since the two negative peaks have a similar size, it also shows that the silanization does not significantly reduce the oxide growth on low-stress \ce{SiN_x}.
	
	\begin{figure}[!t]
		\includegraphics[width=\columnwidth]{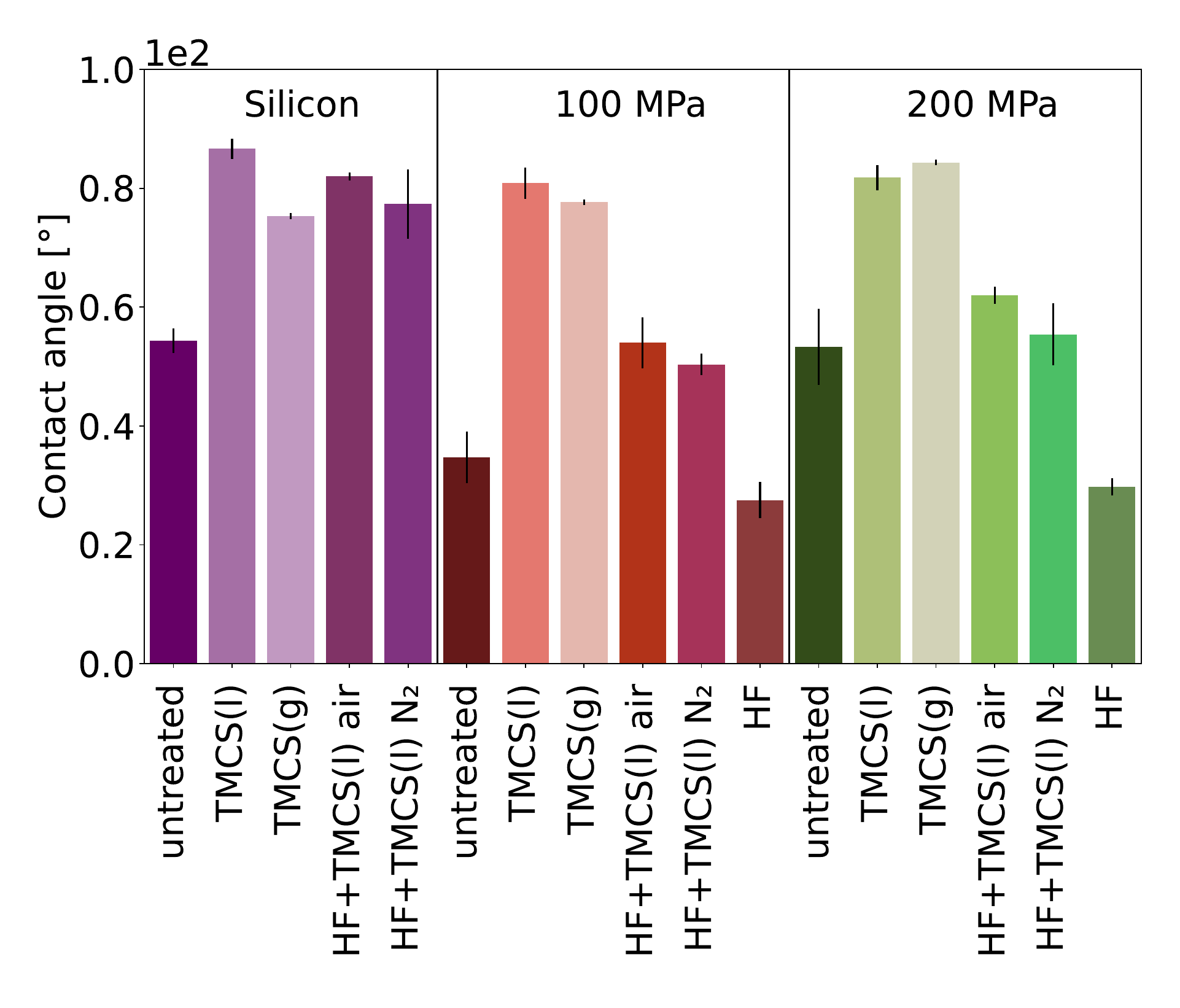}
		\caption{The Water Contact Angles for \ce{SiN_x} samples and silicon before and after several treatments. Silicon is included as a reference.}
		\label{fig:CA}
	\end{figure}
	
	\autoref{fig:CA} summarizes the static water Contact Angle (CA) for both \ce{SiN_x} samples subjected to various surface treatments.
	
	A silicon substrate was measured as a reference for the CA. The increase in CA can be directly correlated with the surface coverage by the hydrophobic \ce{Si-(CH3)3} groups. It shows that all treatments with TMCS increase the CA, while the gas-phase treatment appears to result in a smaller increase than the liquid-phase treatment, when there was no HF etching of the oxide layer before the silanization. This may be due to the fact that in liquid, the TMCS is in dissolved in ethanol, which may remove surface adsorbates like adventitious carbon from the surface, increasing the efficiency of silanization. With HF+TMCS, the CA is somewhat in between, indicating that there may be fewer hydroxyl groups available for the silanization reaction (see \autoref{fig:HF}).\cite{Liu2016}
	
	The untreated 100\,MPa sample has a lower CA than the silicon surface, which has been discussed above. After TMCS treatment without HF etching, the increase in CA shows similar behavior to that on the silicon substrate. However, if HF etching is performed before silanization, the CA does not reach comparable values, indicating that the surface functional groups on the \ce{SiN_x} substrate after HF etching are not sufficient for good TMCS coverage. If the CA was measured immediately after HF etching, the surface was found to be more hydrophilic, which shows that there are indeed hydrophilic groups on the surface. However, since TMCS coverage is lower after HF etching, the surface is likely not dominated by \ce{Si-OH} but is mainly covered with \ce{Si-F}, which TMCS does not bind to.\cite{Liu2016}
	
	Similar findings have been observed for the 200\,MPa samples, except for the untreated 200\,MPa sample, which shows a higher CA than the untreated 100\,MPa sample.
	
	\begin{figure}[!th]
		\includegraphics[width=\columnwidth]{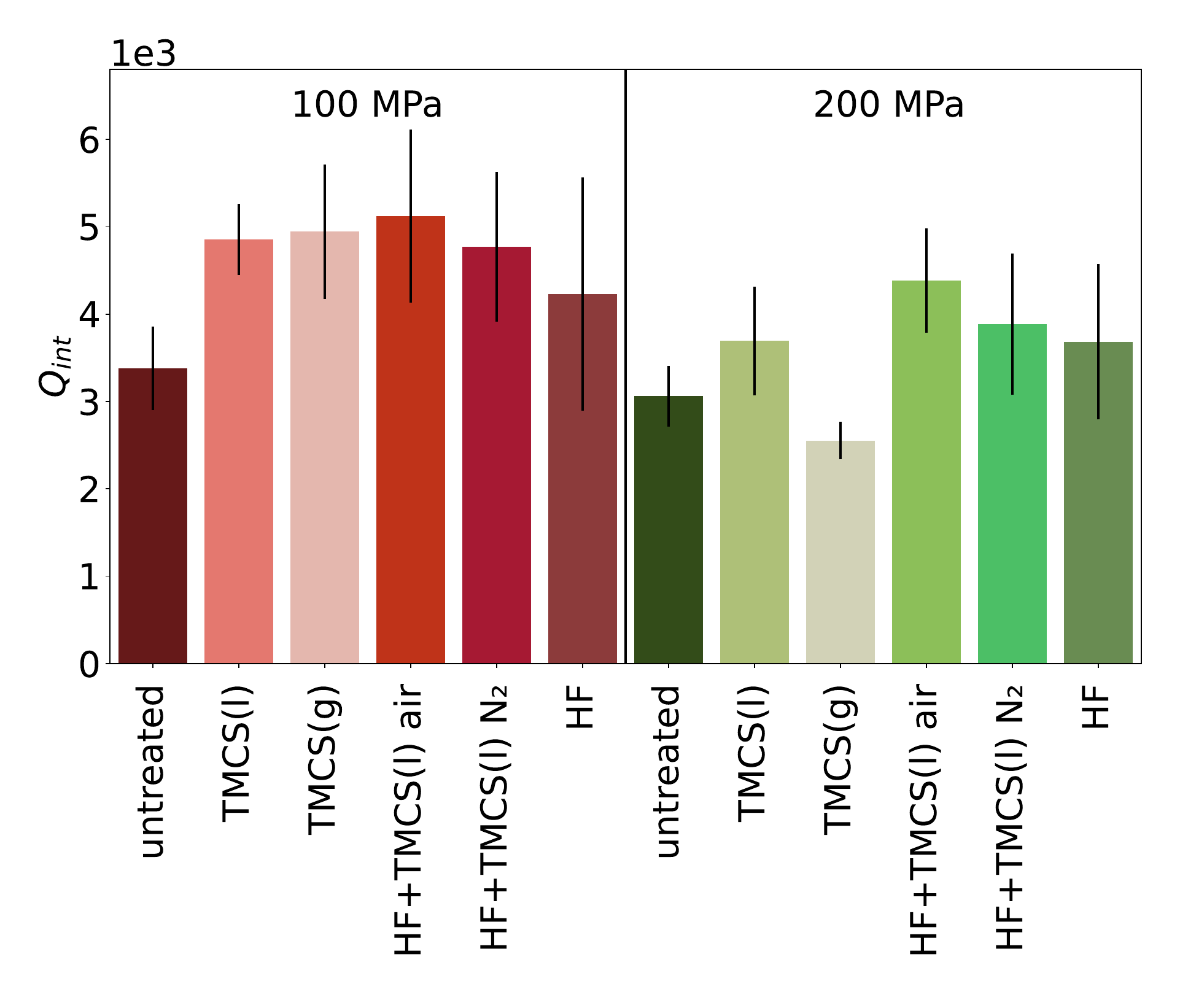}			
		\caption{The $Q_{int}$ values for both \ce{SiN_x} samples before and after several treatments.}
		\label{fig:Q}
	\end{figure}
	
	\begin{figure*}[!ht]
		\includegraphics[width=\textwidth]{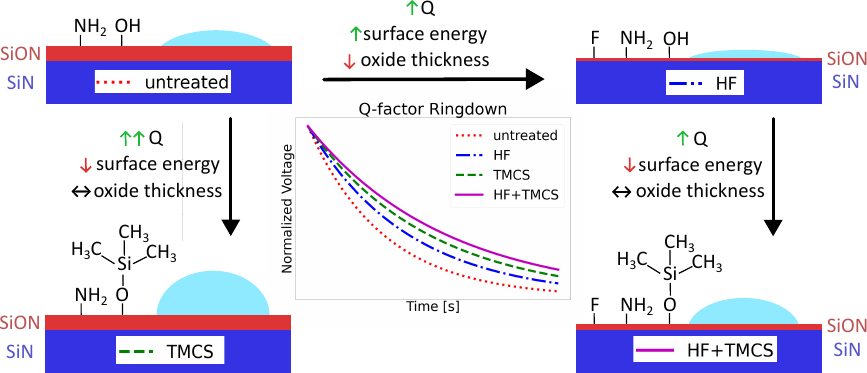}
		\caption{Summary of the surface treatments for the 100\,MPa sample for four parameters: The native oxide layer thickness (red film), the surface energy (water contact angle), proposed surface functional groups and the Q-factor via ring-down plot in the middle.}
		\label{fig:summary}
	\end{figure*}
	
	\autoref{fig:Q} shows the effects of the treatments on the $Q_{int}$. For the 100\,MPa sample, all treatments improved $Q_{int}$, with the TMCS-based approach yielding the largest increases. HF+TMCS(l) air leads to an increase of 50\%, while HF+TMCS(l) \ce{N2} show a more modest increase (40\%) - likely because post-HF surface hydroxyl groups reform more readily upon air exposure. TMCS(l) and TMCS(g) treatments alone yielded comparable enhancements of 45\%, whereas HF etching alone produced the smallest increase at 25\%.
	
	For the 200\,MPa sample, the results show a similar overall trend but with lower improvements: HF+TMCS(l) in air delivered the strongest result at 45\% increase, followed by HF+TMCS(l) in \ce{N2} at 30\%, and TMCS(l) at 20\%. Notably, TMCS(g) treatment resulted in a 20\% decrease in $Q_{int}$, suggesting the untreated 200 MPa surface chemistry is incompatible with gas-phase silanization — a response distinct from the 100 MPa sample. HF etching alone produced a 20\% improvement, similar to the small increase on the 100\,MPa sample. These overall lower gains on the 200 MPa sample likely reflect inherent surface chemistry differences between the samples, with the 200 MPa surface exhibiting a higher contact angle and lower surface energy in its untreated state. 
	
	\autoref{fig:summary} shows a comprehensive summary for the 100\,MPa samples for different treatments, displaying the corresponding water contact angles, which inversely correlate with the surface energy, and the proposed surface functional groups resulting from each treatment.\cite{Han2019,Liu2016} 
	The native oxide layer on \ce{SiN_x} characteristically contains surface hydroxyl and amine groups. Note that the primary amine in the \autoref{fig:summary} is symbolic, representing both primary and secondary surface amines. \cite{Fubini1989,Lamagna2012,Harame1987}
	Both HF etching and silanization substantially reduce the amount of surface hydroxyl groups; HF etching produces predominantly \ce{Si-F} groups, whereas silanization yields \ce{Si-(CH3)3} groups.\cite{Liu2016}\\
	
	\textbf{Native oxide layer} These findings demonstrate that TMCS treatments, with or without native oxide removal, produce substantially greater $Q_{int}$ improvements on low-stress \ce{SiN_x} than HF etching alone. This result clearly shows that surface losses are not attributable to the compressive native oxide layer on low-stress \ce{SiN_x}.
	
	\textbf{Surface energy} Water contact angle measurements, as presented in \autoref{fig:summary} are an indicator of surface energy changes due to treatments, and they have been confirmed by more thorough surface energy determination (SI Table 2). Surface energy values increase with HF etching, whereas TMCS treatments decrease them, yet both treatments enhance $Q_{int}$. Consequently, surface losses do not correlate with the surface energy.
	
	\textbf{Adsorbates}
	Surface adsorbates, such as airborne impurities and moisture, can modify the surface stress and contribute to surface losses.\cite{Yang2002,Chen2017} 
	However, two observations exclude adsorbates as the primary loss mechanism. First, airborne hydrocarbon preferentially adsorbs on high-energy surfaces, yet HF etching, which increases surface energy, still improves $Q_{int}$. \cite{Shimizu2010,Wu2014,Schlangen1995,DellaCiana2023} Second, moisture uptake correlates with wettability, further indicating that adsorbates are not the main reason for the surface losses.\cite{Chen2018,Muster2001} 
	
	\textbf{Surface hydroxyl groups} The $Q_{int}$ increase correlates with a decrease in surface hydroxyl groups. Literature reports that the hydroxyl groups are dominant on the native surface of \ce{SiN_x}.\cite{Lamagna2012,Pezotti2024} After treatments, hydroxyl coverage decreases to markedly reduced but nonzero levels after HF etching, to essentially zero on TMCS-treated samples, where TMCS chemically binds to and eliminates OH groups.\cite{Liu2016}

	Surface hydroxyl groups likely play a dominant role in surface losses.\cite{Ono2005} While hydroxyl groups interact with airborne moisture and hydrocarbon species, literature shows that adsorbate adsorption correlates with surface energy, and this can therefore not be the dominant loss mechanism. Instead, the removal of hydroxyl groups via TMCS silanization or their partial removal via HF etching correlates directly with $Q_{int}$ improvements, suggesting hydroxyl-mediated dissipation as the underlying mechanism. \cite{Nicola}
	
	\section{Conclusion}
	We investigated how HF etching, TMCS silanization, and their combination affect the intrinsic quality factor $Q_{int}$ of low-stress \ce{SiN_x} membranes at nominal stresses of 100 and 200\,MPa. TMCS treatments yield the largest gains (45--50\%), exceeding those from HF etching alone (20--25\%).
	
	These improvements correlate with neither the native oxide thickness nor the surface energy: HF etching thins the oxide and raises the surface energy, whereas TMCS leaves the oxide intact and lowers the surface energy, yet both enhance $Q_{int}$. Since airborne adsorbate uptake scales with surface energy, this absence of correlation also argues against adsorbates as the primary loss channel. The one surface change common to both treatments is a reduction in surface hydroxyl coverage, via \ce{Si-F} formation after HF etching and \ce{Si-(CH3)3} termination after silanization, which correlates with the $Q_{int}$ improvement and provides evidence for surface hydroxyl groups as the relevant loss-bearing species. 
	
	The underlying dissipation mechanism remains open. Hydrogen bonding or dipole coupling of OH groups to the oscillating strain field, as well as interactions with hydrogen-bonded water molecules, are potential loss channels.

	
	\section{Methods}
	\subsection{Silicon nitride resonators}
	Membranes were fabricated from Si wafers coated with 50\,nm low-stress silicon-rich silicon nitride by low-pressure chemical vapor deposition (LPCVD), purchased from Hahn-Schickard-Gesellschaft für angewandte Forschung e.V. with a nominal tensile stress of 100\,MPa and 200\,MPa. The square membranes (length=1\,mm) were patterned by photolithography and dry etching of the backside silicon nitride layer and subsequently released by anisotropic KOH (40\,wt.\%) wet etching all through the silicon wafer.
	
	\subsection{Chemical treatments}
	The prepared resonator chips were treated with several chemical treatments; with HF etching or silanization with trimethylchlorosilane (TMCS) or a combinations of those two. The reaction schemes are shown in \autoref{fig:HF} and \autoref{fig:TMCS}. \autoref{fig:HF} shows the HF etching of a \ce{SiN_x} substrate with a native oxide layer, which is presented with terminating surface hydroxyl groups. Via HF etching, the native oxide is removed, and the \ce{SiN_x} is shown with terminating amine, hydroxide, and fluoride groups, which Liu et al. have shown.\cite{Liu2016} \autoref{fig:TMCS} shows the silanization with TMCS; surface hydroxyl groups are needed for TMCS to bind to the surface, while HCl is released.
	\begin{itemize}
		\begin{figure}[ht]
			\centering
			\includegraphics[width=.8\columnwidth]{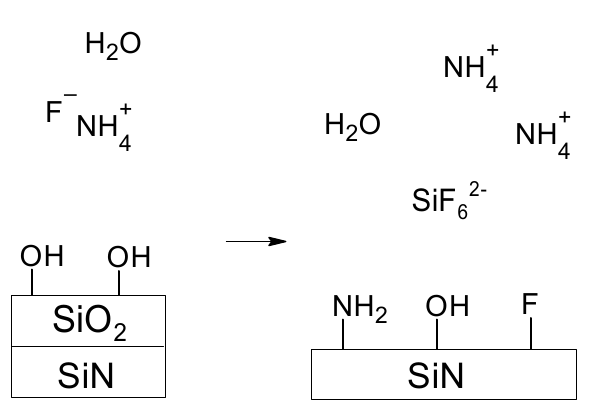}			
			\caption{Simplified scheme. \ce{SiN_x} surface before and after etching with buffered HF (\ce{NH4F})}
			\label{fig:HF}
		\end{figure}
		\item \textbf{HF:} Samples were etched in a 10\% buffered HF solution (\ce{NH4F}) for 10 seconds, washed twice in DI water for 4 seconds each, dipped into ethanol abs. for quicker drying and to avoid water coffee ring stains, and then dried and kept in an \ce{N2} atmosphere until measurement.
		\begin{figure}[ht]
			\centering
			\includegraphics[width=.8\columnwidth]{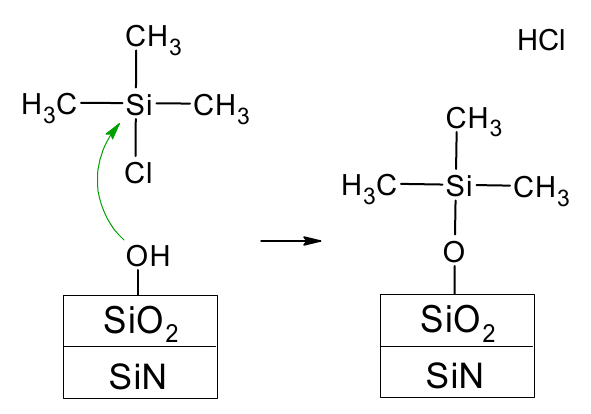}
			\caption{Simplified scheme. \ce{SiN_x} surface before and after TMCS treatment.}
			\label{fig:TMCS}
		\end{figure}
		\item \textbf{TMCS(g):} Samples were placed in a glass chamber under ambient conditions. 1\,mL of Trimethylchlorosilane (TMCS) was put into a glass vial inside of the chamber. The chamber was heated to 75°C for 15 minutes and then let to cool off. 
		\item \textbf{TMCS(l):} Samples were washed with ethanol abs. and then placed into 19\,mL of ethanol abs., into which 1\,mL of TMCS was added slowly while stirring. The reaction took place for 3\, hours. The samples were then washed 3 times with ethanol abs., isopropanol abs. and then dried with a gentle \ce{N2} stream.
		\item \textbf{HF+TMCS(l) air:} HF etching was done as described above, but the sample was dried under ambient conditions instead of \ce{N2}. The silanization reaction was done as described for TMCS(l).
		\item \textbf{HF+TMCS(l) \ce{N2}:} HF etching and TMCS(l) silanization were done as described above.
	\end{itemize}

	\subsection{Characterizations}
	
	\subsubsection{Q and stress measurements}
	Vibrational analysis was done in high vacuum ($p<1\cdot10^{-5}$mbar) with a laser-Doppler vibrometer (MSA-500 from Polytec GmbH). The quality factor Q measurements were done with ring-down measurements with a lock-in amplifier (MFLI from Zurich Instrument).   The intrinsic quality factor $Q_{int}$ was determined by fitting the following equation to the upper envelope of measured $Q$'s as a function of $n^2+j^2$: 
	\begin{equation}
		Q_{int} = \left[\frac{\pi^2(n^2+j^2)}{12}\frac{E}{\sigma}\frac{h^2}{L^2} + \frac{h}{L}\sqrt{\frac{E}{3\sigma}} \right]\cdot Q
		\label{Eq:Qint}
	\end{equation}
	with $n$ and $j$ being the mode numbers; $E$, the Young's modulus, which throughout this work was assumed to be 250\,GPa, independent on stress and frequency; $h$ is the resonator's thickness of 50\,nm; $L$ is the length of the resonator, 1\,mm; $\sigma$ is the stress, which was determined for each measurement, by evaluating a linear regression according to:
	\begin{equation}
		f_{n,m} = \frac{\sqrt{n^2+j^2}}{2L}\sqrt{\frac{\sigma}{\rho}}
		\label{Eq:stress}
	\end{equation}
	with the density $\rho$ taken from the XRR results. (\autoref{tab:prelim}). For each sample, several modes have been measured (up to $n^2+j^2=100$). An example image can be found in the SI. The reported $Q_{int}$ is the average value from the three highest Q factor values in the set.
	
	\subsubsection{Contact Angle and Surface Energy}
	Contact Angle measurements were done on a Kruss DSA30R drop shape analyzer. All samples were cleaned with Ethanol abs. and blow dried with \ce{N2} before measurement to get rid of carbon impurities and adsorbed water. 2 $\mu$l of \ce{H2O} DI or diiodomethane was placed on top of the chip and the CA was measured for 120 seconds. The average values were taken. Contact angle (CA) results given in the main work are for water only. Diiodomethane was used to determine the surface energy of the surfaces, the values are shown in the SI.
	
	The surface energy was determined with the OWRK/Fowkes approach:\cite{Owens1969}
	\begin{equation}
		0.5\cdot \gamma_{LG}\cdot(1+cos\theta) = \sqrt{\gamma_{SG}^D\cdot\gamma_{LG}^D}+\sqrt{\gamma_{SG}^P\cdot\gamma_{LG}^P}
	\end{equation}
	where $\gamma$ is the interfacial tension, more specifically, $\gamma_{LG}$ the surface tension of the liquid with the gas space, which is the sum of dispersive $\gamma_{LG}^D$ and polar $\gamma_{LG}^P$ components, and they are known for the used substances water and diiodomethane. The unknown variables in this equation are $\gamma_{SG} = \gamma_{SG}^D+\gamma_{SG}^P$, which is the interfacial tension of the solid with the gas, i.e., the surface energy. 
	
	\subsubsection{AFM}
	The surface topology was analyzed using ScanAsyst equipment (Bruker, Dimension Edge with Scan Asyst). An antimony-doped silicon tip (Value AFM Probes by Bruker, $T = 4\ \mu m$, $k = 42\ N/m$, $f_0 = 320\ kHz$) with reflective aluminium on the back side was used in tapping mode. Each measurement was performed on the sample frame, which had been cleaned with isopropanol and a gentle \ce{N2} stream. Images were recorded using height, phase, and amplitude channels of square frames of $1\mu m\cdot1\mu m$ with a resolution of 512\,nm in both x and y direction and a scanning rate of $0.5\mu m/s$. The root-mean-squared (RMS) surface roughness was evaluated using the Python surfalize package. \cite{Schell2024}

	\subsubsection{RBS \& ERDA}
	Samples have been measured with 2 MeV He for RBS and 13 MeV $^{127}I$ Heavy Ion ToF-ERDA. Elemental Ratios are given in atomic fractions. For RBS, a SIMNRA simulation was carried out and is depicted in the plot as a blue solid line. The areal density is given in $atoms/cm^2$. The uncertainty of the given area density is approximately 5\%. The ERDA spectra were analyzed with the Potku software, and depth profiles were created. The elemental composition was determined only in the depth region between the blue and red markers. The depth scale is $atoms/cm^2$ and depth profiles have been created for H, N, O, Si, and Cl. The O accumulation is recognizable at the surface and interface (O double peak). The errors of the elemental ratios are estimated to be of the order of 7\% and 10\% for H, unless otherwise stated. Since the atomic fraction results of RBS and ERDA fit together, the results of ERDA are depicted. The depth in $atcm^2$ was converted to nm with the density of \ce{SiN_x} results from XRR and the molecular mass according to the ERDA ratio, like this:
	\begin{equation}
		d[nm] = \frac{D[at./cm^2]}{\frac{\rho[g/cm^3]\cdot N_A[1/mol]}{M[g/mol]}}\cdot10^7[nm/cm]
	\end{equation}
	where D is the atomic depth as measured, $\rho$ the density taken form XRR, $N_A$ the Avogadro constant and M the molecular mass according to the Si/N ratio, taken from the ERDA measurement.

	The error of the depth resolution is $\approx$10\,nm.
	
	\subsubsection{FTIR}
	FTIR spectra were taken from 400-4000 cm\textsuperscript{-1} and a resolution of 4~cm$^{-1}$ with nanoelectromechanical photothermal Fourier-transform infrared spectroscopy (NEMS-FTIR): A chamber holding the membrane in 10\textsuperscript{-5}~mbar, with the membrane optically driven and read-out with a 633\,nm laser in conjunction with an FTIR spectrometer (Vertex 70, Bruker Optics, MA, USA). More details regarding the chamber can be found elsewhere.\cite{TimaracPopovic2026}
	
	\subsubsection{TEM-EELS}
	The samples were prepared for cross-sectional TEM measurements by cutting, mechanical grinding, polishing, and ion milling. 
	For the TEM and STEM-EELS investigations, a FEI TECNAI G20 TEM (with FEG emittant, Gatan Orius 600 camera and a Gatan GIF 2001 energy filter, STEM point resolution 0.2\,nm) and a FEI TECNAI F20 (with X-FEG emittant, Gatan Rio16, 30 fps full HD camera and a Gatan GIF Tridiem energy filter, STEM point resolution 0.15\,nm) were used for the 100\,MPa and the 200\,MPa sample, respectively.
	The parts of the cross-sectional sample measured during STEM line profiles were between 60 and 90\,nm thick. During EELS profiling, the dwell time was 1\,s and 250 points were measured over a length of 70-85\,nm. The line profile resolution was determined to be around 0.6-0.7\,nm. Before the measurements, the samples were plasma cleaned using a Helium plasma for 30\,s.

	\subsubsection{XPS}
	XPS measurements were performed with a ULVAC-PHI Quantera SXM and a PHI Versaprobe III, both equipped with a monochromatic Al K$\alpha$ source (Excitation energy: 1486.6\,eV; Beam energy and spot size: 50\,W onto 200\,$\mu m$). Pass energies of 140\,eV and 27\,eV and step widths of 0.25 and 0.05\,eV were used for survey and detail spectra, respectively. Argon sputtering was done on an area of $2\cdot 2mm^2$ with 2\,kV for 5\, minutes and measured at an electron take-off angle of 45°, to remove the surface oxide layer and determine the bulk N/Si ratio. Angle-resolved measurements were done at electron take-off angles between 45 and 90° to the sample surface to determine the oxide thickness. 
	
	Data analysis was done with MultiPak and CasaXPS.	
	Thickness determination was done with the minimization of \autoref{Eq:scipy} from Seah et al.\cite{Seah2002} with the python function scipy.optimize.minimize\_scalar:\\
	\begin{equation}
		R = \frac{I_{layer}}{I_{substrate}} = R^{inf}\cdot\frac{1-exp[-d_{layer}/\lambda_{O\,1s,layer} \cdot sin{\theta}]}{exp[-d_{layer}/\lambda_{Si\,2p,layer} \cdot sin{\theta}]}
		\label{Eq:scipy}
	\end{equation}
	where $I_{layer}$ is the signal from the layer and $I_{substrate}$ the signal from the substrate; $d_{layer}$ is the thickness of the layer; $\lambda_{shell,material}$ is the inelastic mean free path (IMFP) of the electron coming from a shell in the material; $R^{inf}$ is the ratio of the layer and substrate materials at infinite thicknesses, and is calculated by:\cite{Hesse2005} 
	
	\begin{equation}
		R^{inf}=\frac{I^{inf}_{layer}}{I^{inf}_{substrate}}=\frac{N_{O\,1s,layer}\cdot \lambda_{O\,1s,layer}}{N_{Si\,2p,substrate}\cdot \lambda_{Si\,2p,substrate}}
		\label{Eq:Rinf}
	\end{equation}
	where $N=N_A \rho /A$ is the atomic density: $N_A$ is Avogadro's constant, $\rho$ is the density of the material and A is the relative atomic mass (keep in mind, that the relative atomic mass needs to be normalized, e.g. A(\ce{SiO_2}) needs to be divided by 2, to give the formula wight for 1 mole of O atoms). $\lambda_{shell,origin}$ is the inelastic mean free path (IMFP) of an electron from a specific shell in the layer/bulk.
	
	The values taken for the evaluation were partly taken from XRR measurements in this work, such as $\rho=2.95\ g/cm^3$ for 100 \,MPa and $\rho=3\ g/cm^3$ for 200\,MPa \ce{SiN_x}, and partly taken from literature, such as $\rho=2.33\ g/cm^3$ for \ce{Si}, $\rho=1.5\ g/cm^3$ for \ce{SiO_2} when evaluating the thickness on \ce{SiN_x}\cite{Antonius}, while $\rho=2.2\ g/cm^3$\cite{Seah2002} was taken when evaluation the thickness of the native oxide on plain \ce{Si}. The IMFP was taken from QUASES, an IMFP calculation by the TPP2M formula.\cite{Tanuma1994}
	
	Evaluated thicknesses were compared with StrataPHI and SESSA evaluation to verify this method.		 
	
	\subsubsection{XRR}
	X-ray reflectometry (XRR) was performed on a Bruker D8 Discover Plus diffractometer, $\theta$--$\theta$ geometry) equipped with a TXS rotating-anode \ce{Cu} source operated at 45~kV and 120~mA (\ce{Cu} K$\alpha$, $\lambda = 1.5406$~\AA). The primary beam was conditioned by a focusing G\"obel mirror, a 0.05~mm exit slit, and a $22 \times 1.8$~mm collimator; the reflected beam passed a motorized anti-scatter slit and was recorded on an Eiger2 R 500K detector in 0D mode (goniometer radius 380~mm). After manual alignment of the sample height, tilt ($\omega$), and azimuth ($\chi$) to the direct beam, reflectivity was acquired as a coupled $\theta$--$2\theta$ scan from $2\theta = 0.03$ to $7.5^\circ$ in steps of $0.005^\circ$ with 3~s per step. Data analysis was conducted using GenX software \cite{Glavic2022GenX3}.

	\section*{Acknowledgements}
	The authors thank the Analytical Instrumentation Center of TU Wien for the XPS measurements, which were carried out using infrastructure funded by the FFG “ELSA” project under grant No. 884672. We further acknowledge Juergen Smoliner, Alois Lugstein, and Masiar Sistani from the Institute of Solid State Electronics at TU Wien for granting us access to their FTIR spectrometer. This project received funding from the Austrian Science Fund (FWF) under project 10.55776/I6086.
	

	\bibliographystyle{elsarticle-num}
	\bibliography{arilib}
	
	
	
	

	\clearpage
	\onecolumn
	\begin{center}
		\textbf{\large Supplementary Information}
	\end{center}
	\vspace{1cm}
	
	\setcounter{section}{0}
	\setcounter{figure}{0}
	\setcounter{equation}{0}
	\setcounter{table}{0}
	\renewcommand{\thesection}{S\arabic{section}}
	\renewcommand{\thefigure}{S\arabic{figure}}
	\renewcommand{\theequation}{S\arabic{equation}}
	\renewcommand{\thetable}{S\arabic{table}}
	\section{Q and $Q_{int}$ plot}
	
\begin{figure}[h]
		\includegraphics[width=\textwidth]{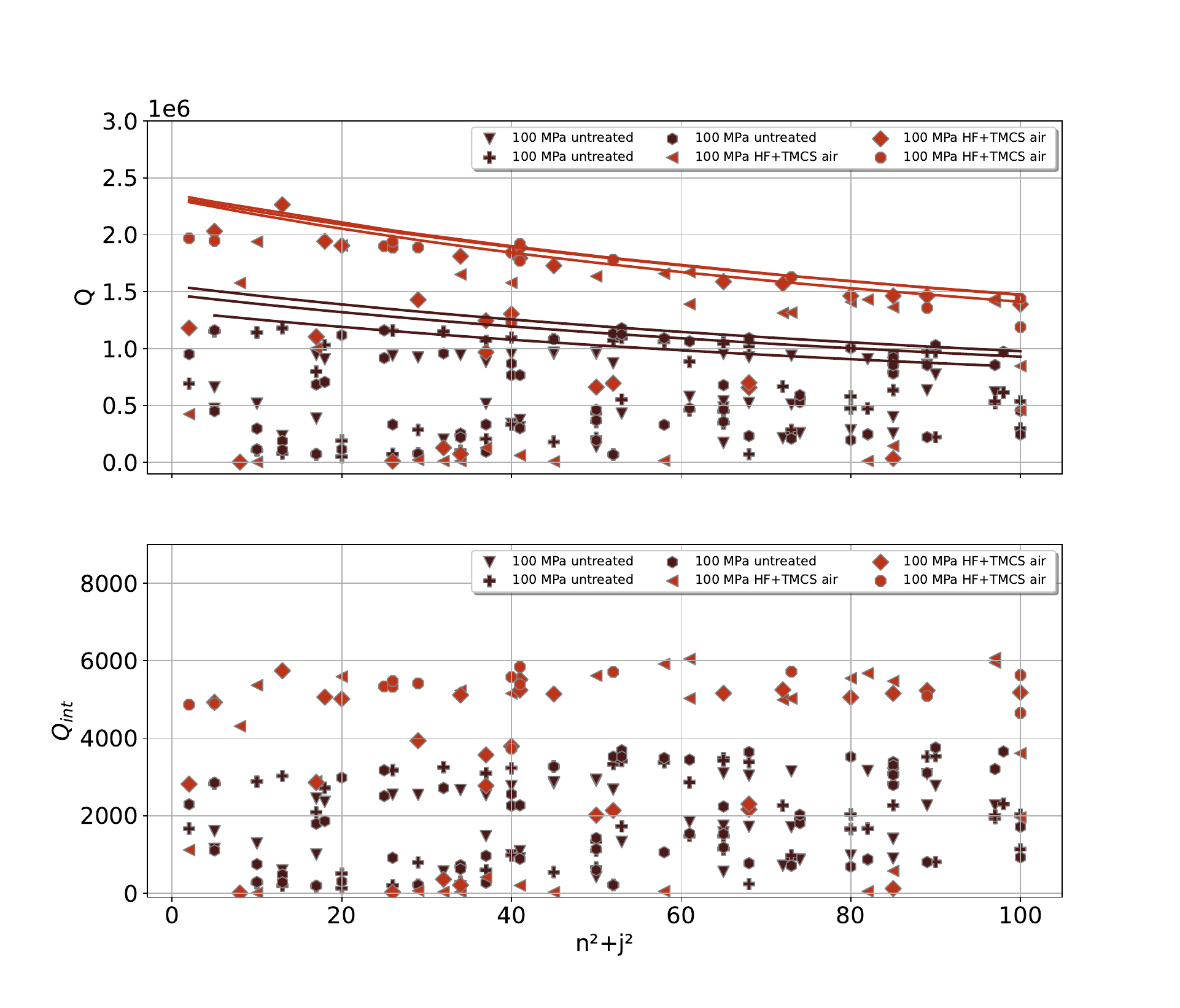}
		\caption{Q factor (top) and determined $Q_{int}$ (bottom) for modes $n^2+j^2 \le 100$. The lines in the top plot represent the highest quality factor value envelope due to intrinsic loss.}
		\label{fig:Qplot}
	\end{figure}

	\autoref{fig:Qplot} shows the Qs and intrinsic Qs ($Q_{int}$) for modes $n^2+j^2 \le 100$. The lines in the top plot represent the envelope from the mean of the three largest Q values due to intrinsic loss according to \autoref{Eq:Qint}.
	\begin{equation}
		Q_{int} = \left[\frac{\pi^2(n^2+j^2)}{12}\frac{E}{\sigma}\frac{h^2}{L^2} + \frac{h}{L}\sqrt{\frac{E}{3\sigma}} \right]\cdot Q
		\label{Eq:Qint}
	\end{equation}
	
	\newpage
	\section{Surface tension}\label{sec:ST}
	Surface tension was calculated according to \autoref{Eq:ST} according to Ref.\cite{Owens1969}.
	
	\begin{equation}
		\gamma_{LG}\cdot (1+cos(\theta)) = \sqrt{\gamma_{SG}^D \cdot \gamma_{LG}^D} + \sqrt{\gamma_{SG}^P \cdot \gamma_{LG}^P}
		\label{Eq:ST}
	\end{equation}
	with $\theta$ being the contact angle of the respective liquid on the sample surface, $\gamma_{LG}$ the combined surface energy between liquid and gas phase, $\gamma_{SG}$ the combined surface energy between solid (sample) and gas phase, and the superscript $^D$ and $^P$ stand for dispersive and polar fraction, respectively. the used values for $\gamma_{LG}$ can be found in \autoref{tab:tension} and the measured angles and resulting $\gamma_{SG}$ values are displayed in \autoref{tab:SE}.
	
	\begin{table}[h]
		\centering
		\caption{Surface tension (liquid-gas) values for used contact angle liquids; combined values and split into dispersive and polar fractions.\cite{Owens1969}}
		\label{tab:tension}
		\begin{tabular}{lccc} 
			\hline
			Contact angle liquid    &	$\gamma_{LG}\ [mN/m]$     &	$\gamma_{LG}^D\ [mN/m]$	   &	$\gamma_{LG}^P\ [mN/m]$       \\
			\hline
			DI \ce{H2O}		        &	      72.8		          &		21.8		           &         51                      \\
			Diiodomethane	(DIM)   &		50.8		          &		49.5		           &		1.3		                  \\
			\hline
		\end{tabular}
	\end{table}
	
	\begin{table}[h]
		\centering
		\caption{Measured contact angles and determined surface energies of different samples.}
		\label{tab:SE}
		\begin{tabular}{lccccc} 
			\hline
			Sample								&$\theta_{\ce{H2O}}[^{\circ}]$&  $\theta_{DIM}[^{\circ}]$   &	$\gamma_{SG}\ [mN/m]$   &	$\gamma_{SG}^D\ [mN/m]$	&	$\gamma_{SG}^P\ [mN/m]$	\\
			\hline
			100\,MPa							&					&					&					&					&                    \\
			\hspace{.5cm}untreated				&		35      	&		48      	&		61.2		&		25.6        &       35.6		 \\
			\hspace{.5cm}HF      				&		28   		&		32     		&		67.1		&		32.8		&       34.3           \\
			\hspace{.5cm}TMCS                   &	     78         &	     54        	&	     35.3       &	      28.0 	    &        7.3           \\
			\hspace{.5cm}HF+TMCS                &	     54       	&	     45         &	      50.1      &	     29.5     	&        20.7            \\
			200\,MPa							&					&					&					&					&                    \\
			\hspace{.5cm}untreated				&		53      	&		48      	&		50      	&		27.7        &       22.3		 \\
			\hspace{.5cm}HF      				&		30   		&		36     		&		65.5       	&		31.1		&       34.4           \\
			\hspace{.5cm}TMCS                   &	     83         &	     54        	&	     33.8       &	      28.9 	    &        4.9           \\
			\hspace{.5cm}HF+TMCS                &	     62       	&	     48         &	     44.7       &	     29.0     	&        15.8            \\
			\hline
		\end{tabular}
	\end{table}
	
	\newpage
	\section{Depth profiling 100\,MPa sample}
	
	\begin{figure}[h!]
		\centering
		\begin{subfigure}[b]{0.49\textwidth}
			\centering
			\includegraphics[width=\textwidth]{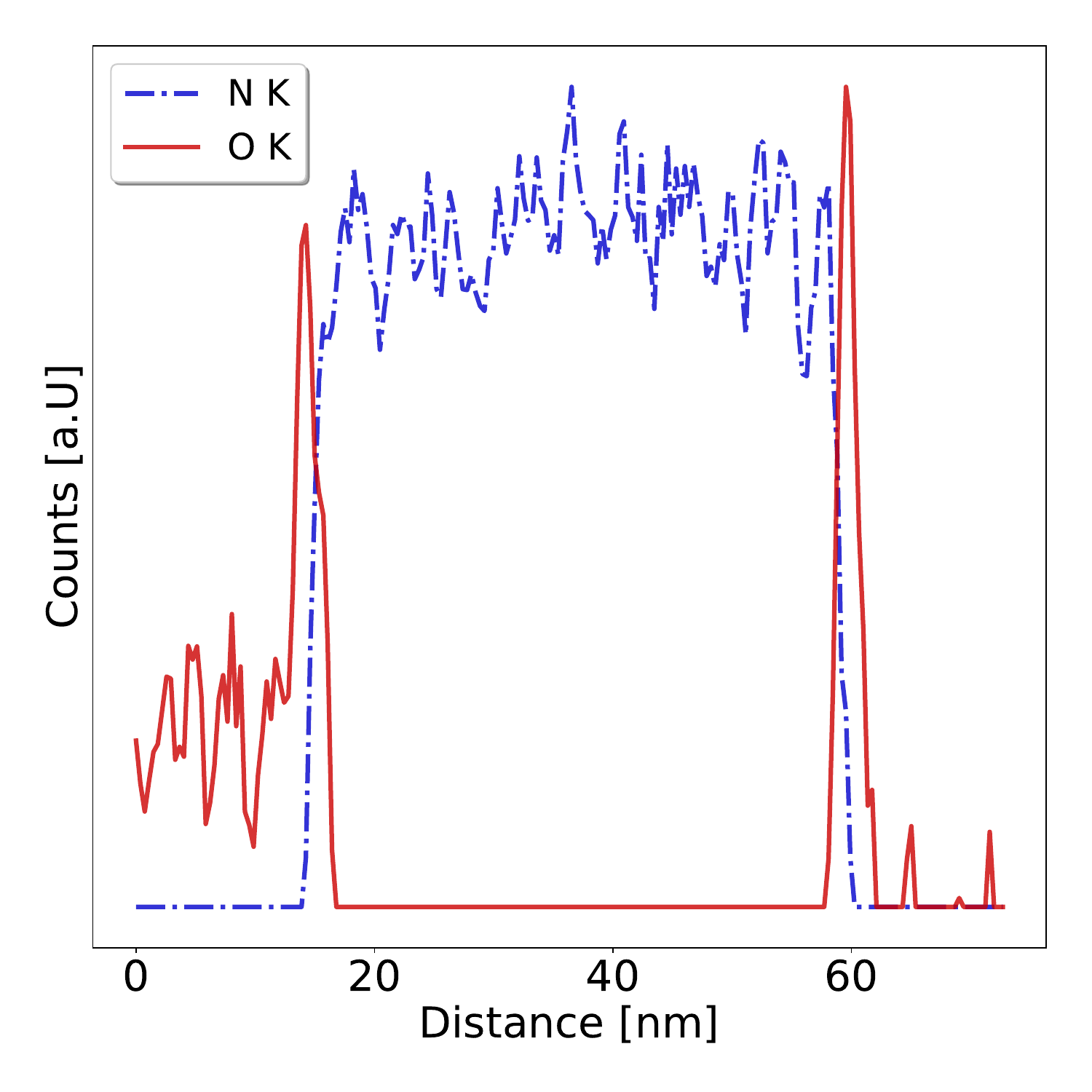}
			\caption{}
			\label{fig:100MPaEELS}
		\end{subfigure}%
		\hfill
		\begin{subfigure}[b]{0.49\textwidth}
			\centering
			\includegraphics[width=\textwidth]{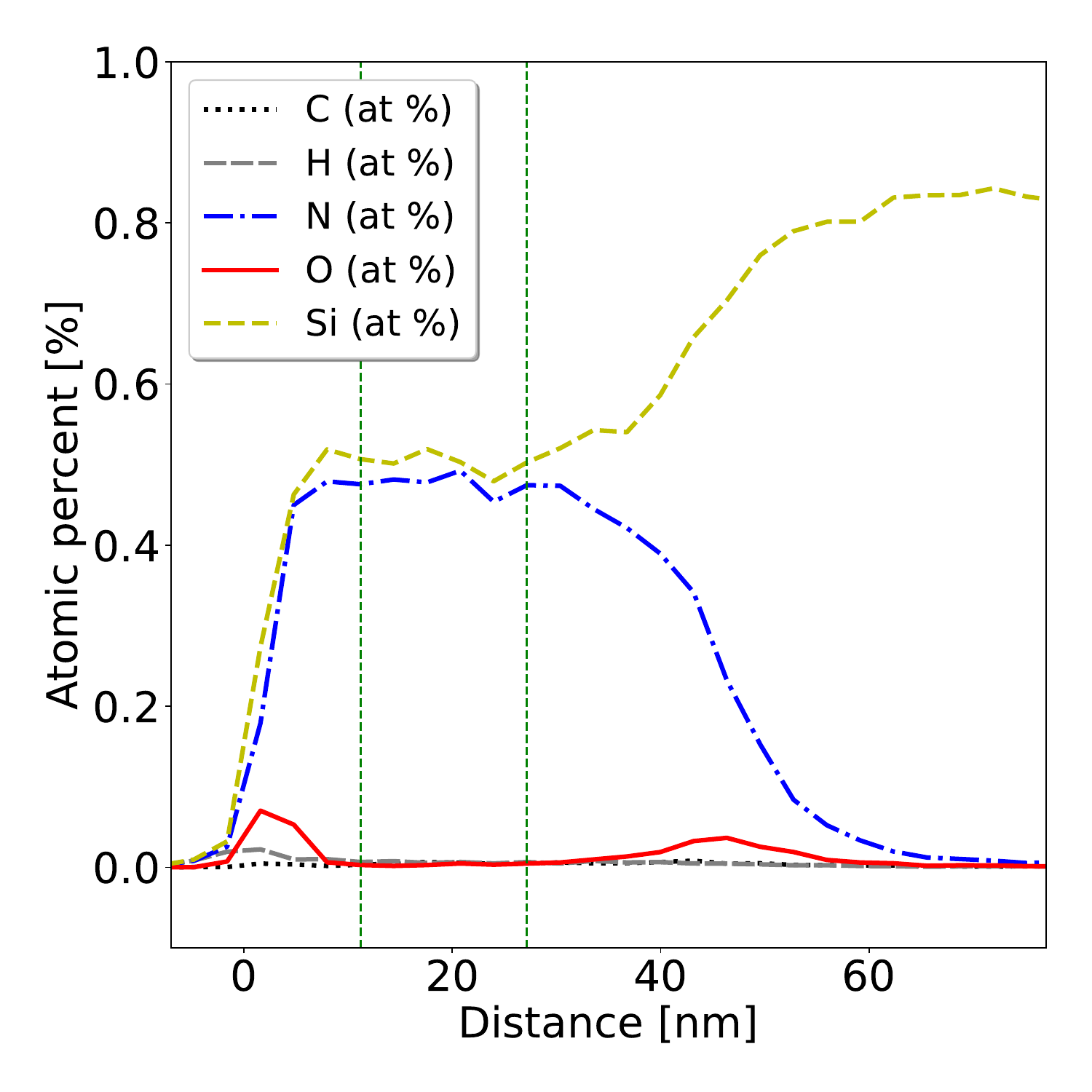}
			\caption{}
			\label{fig:100MPaERDA}
		\end{subfigure}%
		\caption{Depth profiling methods for the 100\,MPa sample. a)  EELS line scan analysis across the multilayers; b) ERDA depth profile}
		\label{fig:crosssection2}
	\end{figure}
\end{document}